\documentclass[prb,10pt,twocolumn]{revtex4}

\usepackage{epsfig}
\graphicspath{{/}}

\begin{document}

\title{Radiative and Non-Radiative Exciton Energy Transfer in Monolayers of Two-Dimensional Transition Metal Dichalcogenides}

\author{Christina Manolatou, Haining Wang, Weimin Chan, Sandip Tiwari, Farhan Rana}
\affiliation{School of Electrical and Computer Engineering, Cornell University, Ithaca, NY 14853}
\email{fr37@cornell.edu}

\begin{abstract}
  We present results on the rates of interlayer energy transfer between excitons in two-dimensional transition metal dichalcogenides (TMDs). We consider both radiative (mediated by real photons) and non-radiative (mediated by virtual photons) mechanisms of energy transfer using a unified Green's function approach that takes into account modification of the exciton energy dispersions as a result of interactions. The large optical oscillator strengths associated with excitons in TMDs result in very fast energy transfer rates. The energy transfer times depend on the exciton momentum, exciton linewidth, and the interlayer separation and can range from values less than 100 femtoseconds to more than tens of picoseconds. Whereas inside the light cone the energy transfer rates of longitudinal and transverse excitons are comparable, outside the light cone the energy transfer rates of longitudinal excitons far exceed those of transverse excitons. Average energy transfer times for a thermal ensemble of longitudinal and transverse excitons is temperature dependent and can be smaller than a picosecond at room temperature for interlayer separations smaller than 10 nm. Energy transfer times of localized excitons range from values less than a picosecond to several tens of picoseconds. When the exciton scattering and dephasing rates are small, energy transfer dynamics exhibit coherent oscillations. Our results show that electromagnetic interlayer energy transfer can be an efficient mechanism for energy exchange between TMD monolayers.     
\end{abstract}
                                    
\maketitle

\section{Introduction}
The optoelectronic properties of 2D transition metal dichalcogenide (TMD) monolayers are dominated by excitons~\cite{fai10,fai12,xu13}. Distinguishing features of the excitons in 2D metal dichalcogenides are the large exciton binding energies and the strong exciton-photon interactions~\cite{fai10,fai12,xu13,Changjian14,timothy,Chernikov14,Wang16}. Recently, exciton-polaritons have been also studied experimentally and theoretically in these materials~\cite{Menon14,Vasil15,Wang16,Gartstein15}. The strong exciton-photon coupling results in spontaneous emission radiative lifetimes in the hundreds of femtoseconds range~\cite{Moody15,Huber15,Marie15,Wang16}. The strong exciton-photon coupling suggests that the rates for interlayer energy transfer between excitons in parallel TMD monolayers would also be fast.  

\begin{figure}
  \begin{center}
   \epsfig{file=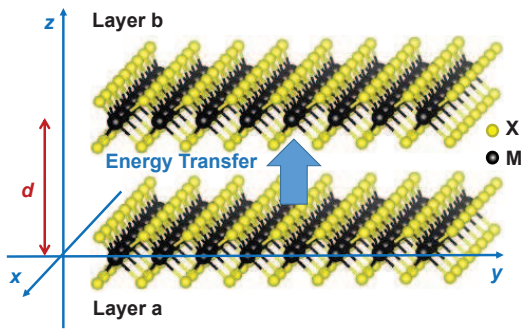,angle=0,width=0.4\textwidth}
    \caption{Exciton energy transfer between two parallel TMD layers separated by a distance $d$ along the $z$-axis.}
    \label{fig:fig1}
  \end{center}
\end{figure}

In electronically coupled 2D TMD monolayers, ultrafast energy transfer via interlayer charge transfer has been observed~\cite{Javey14,Hong14,Rigosi15}. In this work, we study the rate of transfer of energy between excitons in parallel TMD monolayers as a result of electromagnetic coupling. The mechanism for this energy exchange could be both radiative (mediated by propagating photons for exciton states within the light cone~\cite{Wang16}) or non-radiative (mediated by evanescent photons that are bound to the exciton states outside the light cone in an isolated TMD layer as exciton-polaritons but can mediate energy exchange between two TMD layers if the two TMD layers are close enough). The latter mechanism is the same as the well known Forster resonance energy transfer (FRET) mechanism due to dipole-dipole coupling~\cite{Andrews04}. However, the use of the standard FRET dipole-dipole energy exchange formulas in the present case gives erroneous results since it ignores the retarded nature of the exciton interlayer interaction~\cite{Tomita96,Lyo00}. The quantum electrodynamic Green's function approach used here treats both these mechanisms on equal footing while taking into account the corrections to the longitudinal and the transverse exciton energy dispersion relations due to coupling with the radiation. 

The energy transfer times depend on the exciton momentum, exciton intralayer scattering rates, and the interlayer separation. Our results show that the large exciton optical oscillator strengths in TMD monolayers result in energy transfer times shorter than 100 fs for longitudinal excitons for interlayer spacings smaller than 10 nm. Average energy transfer times for a thermal ensemble of longitudinal and transverse excitons can be smaller than a picosecond. We also consider localized excitons and find that localized longitudinal excitons can also have energy transfer times shorter than a picosecond for interlayer spacings smaller than 10 nm. If the exciton scattering rates are fast, energy transfer involves a simple decay of energy from one layer to the other. If the exciton scattering rates are slow, energy transfer dynamics can exhibit coherent oscillations. Conditions for observing such oscillations are discussed in this paper. Our results show that electromagnetic interlayer energy transfer can be an efficient mechanism for energy exchange between electronically uncoupled TMD monolayers and can even compete with the interlayer charge transfer mechanism in the case of electronically coupled TMD layers. 

Section~\ref{sec:prelim} discusses exciton states in TMDs, longitudinal and transverse excitons in TMDs, and the Hamiltonian terms that describe exciton-photon interaction in TMDs. Sections ~\ref{sec:rate} and ~\ref{sec:dis}  present the main results of this paper on the energy transfer rates between two TMD layers. Section~\ref{sec:num} presents numerical results for energy transfer rates between two MoS$_{2}$ layers. Section ~\ref{sec:coh} considers the case where the exciton intralayer scattering and dephasing rates are small and energy transfer dynamics exhibit coherent oscillations. Sections ~\ref{sec:app1} and \ref{sec:app2} discuss some special cases, including that of localized excitons and energy transfer between excitons and free electron-hole pairs in TMD heterolayers.  The Appendix discusses exciton self-energies and energy dispersions in TMD monolayers in the presence of exciton-photon interaction.  

Although most of the numerical results presented in this paper are for MoS$_{2}$ monolayers, the analysis and the results presented are expected to be relevant to all TMDs, and are expected to be useful in realizing new kinds of metal dichalcogenide optoelectronic devices based on energy transfer.

\section{Preliminaries} \label{sec:prelim}

\subsection{Energy Bands in TMDs}
The conduction and valence bands in monolayer TMDs near the $K$ and $K'$ points in the Brillouin zone are described by the Hamiltonian~\cite{Wang16},
\begin{equation}
\left[
\begin{array}{cc}
\Delta/2 & \hbar v k_{-} \\
\hbar v k_{+} & -\Delta/2 + \lambda \tau \sigma
\end{array} \right]    \label{eq:H1}
\end{equation}
Here, $\Delta$ is related to the material bandgap, $\sigma=\pm1$ stands for the electron spin, $\tau=\pm1$ stands for the $K$ and $K'$ valleys, $2\lambda$ is the splitting of the valence band due to spin-orbit coupling, $k_{\pm}=\tau k_{x}\pm ik_{y}$, and the velocity parameter $v$ is related to the coupling between the orbitals on neighboring $M$ atoms. From density functional theories \cite{Lam12,Falko13}, $v\approx 5-6 \times 10^5 $ m/s. The wavevectors are measured from the $K$($K'$) points. The d-orbital basis used in writing the above Hamiltonian are $| d_{z^{2}} \rangle$ and $(| d_{x^{2}-y^{2}}\rangle + i\tau | d_{xy}\rangle)/\sqrt{2}$~\cite{yao12}. We will use the symbol $s$ for the combined valley ($\tau$) and spin ($\sigma$) degrees of freedom. The intravalley momentum matrix element between the conduction and valence band states near $K$($K'$) points follows from the above Hamiltonian,
\begin{eqnarray}
  & & \vec{P}_{vc,s}(\vec{k}',\vec{k}) =  \langle v_{v,\vec{k}',s}| \hat{\vec{P}} | v_{c,\vec{k},s}\rangle \nonumber \\
& & = m_{o} v \, \left[ (\hat{x} + i\tau \hat{y}) \cos(\theta_{\vec{k}',s}/2) \cos(\theta_{\vec{k},s}/2) e^{-i\tau (\phi_{\vec{k}'} + \phi_{\vec{k}})/2} \right. \nonumber \\ 
& & \left. - (\hat{x} - i\tau \hat{y}) \sin(\theta_{\vec{k}',s}/2) \sin(\theta_{\vec{k},s}/2) e^{i\tau (\phi_{\vec{k}'} + \phi_{\vec{k}})/2} \right] \nonumber \\
  \label{eq:matrix}
\end{eqnarray}
Here, $m_{o}$ is the free electron mass, $\phi_{\vec{k}}$ is the phase of the wavevector $\vec{k}$, and,
\begin{equation}
\cos(\theta_{\vec{k},s}) = \frac{\Delta_{s}}{2 \sqrt{ (\Delta_{s}/2)^{2} + (\hbar v k)^{2}}} \label{eq:cs}
\end{equation}
Near the band extrema, $\cos(\theta_{\vec{k},s}) \rightarrow 1$ and $\vec{P}_{vc,s}(\vec{k}',\vec{k}) \rightarrow m_{o}v \,(\hat{x} + i\tau \hat{y})e^{-i\tau (\phi_{\vec{k}'} + \phi_{\vec{k}})/2}$.

\subsection{Exciton States in TMDs}
Exciton states in TMDs have been discussed in detail in several published works~\cite{Wang15b,Wang16,Changjian14,timothy,Chernikov14}. We assume an undoped intrinsic TMD layer. The creation operator for an exciton state with center-of-mass momentum $\vec{Q}$ is defined as~\cite{Wang15b,Wang16},
\begin{equation}
  B^{\dagger}_{\vec{Q},s,\alpha} =  \frac{1}{\sqrt{A}}\sum_{\vec{k}} \psi_{\vec{Q},s,\alpha}(\vec{k}) c_{\vec{k} + \lambda_{e} \vec{Q},s}^{\dagger} b_{\vec{k}-\lambda_{h}\vec{Q},s} 
\end{equation}
$c_{\vec{k},s}$ and $b_{\vec{k},s}$ are the destruction operators for the conduction band and valence band electron states, respectively, $\psi_{\vec{Q},s,\alpha}(\vec{k})$ is the exciton relative wavefunction, and $A$ is the area of the TMD monolayer. $\lambda_{e} = m_{e}/m_{ex}$, $\lambda_{e} = m_{h}/m_{ex}$, $m_{ex} = m_{e} + m_{h}$, where $m_{e}$, $m_{h}$, and $m_{ex}$ are the effective masses of electrons, holes, and excitons, respectively~\cite{Wang16}. The subscript $\alpha$ corresponds to the different bound exciton levels. Only optically active exciton levels (whose relative wavefunction have a non-zero amplitude at zero relative distance) will be considered here. The exciton Hamiltonian can be written approximately as,
\begin{equation}
  H_{ex} = \sum_{\vec{Q},\alpha,s} E_{ex,\alpha}(\vec{Q}) \,   B^{\dagger}_{\vec{Q},s,\alpha}   B_{\vec{Q},s,\alpha}
\end{equation}
Here, $E_{ex,\alpha}(\vec{Q})$ is the exciton energy measured with respect to the ground state consisting of a filled valence band and an empty conduction band.  

\subsection{Exciton-Photon Interaction in TMDs: Exciton-Polaritons}
The quantized radiation field is~\cite{haugbook},
\begin{eqnarray}
  \vec{A}(\vec{r}) & = & \sum_{\vec{q},j} \sqrt{\frac{\hbar}{2\epsilon_{o} \omega_{q}}} \left[ \hat{n}_{\vec{q},j} a_{\vec{q},j} +  \hat{n}^{*}_{-\vec{q},j} a^{\dagger}_{-\vec{q},j} \right] \frac{e^{i\vec{q}.\vec{r}}}{\sqrt{V}} \nonumber \\
  & = & \sum_{\vec{q},j} \sqrt{\frac{\hbar}{2\epsilon_{o} \omega_{q}}} \vec{R}_{\vec{q},j} \frac{e^{i\vec{q}.\vec{r}}}{\sqrt{V}} \label{eq:field}
\end{eqnarray}
Here, $\hat{n}_{\vec{q},j}$ for $j=1,2$ are the field polarization vectors and $a_{\vec{q},j}$ is the field destruction operator for a mode with wavevector $\vec{q}$ and frequency $\omega_{q}$. We assume a TMD monolayer occupying the $x-y$ plane located at $z$ along the $z$-axis. The interaction between the electrons and photons is given by the Hamiltonian,
\begin{equation}
H_{int} =  \sum_{j,s} H^{+}_{j,s} +  H^{-}_{j,s} 
\end{equation}
where,
\begin{eqnarray}
  & &   H^{-}_{j,s}  =  \frac{e}{m_{o}} \sum_{\vec{q},\vec{k}_{\parallel},\alpha} \sqrt{\frac{\hbar}{2 V A \epsilon_{o} \omega_{q}}} e^{-iq_{z}z} \times \nonumber \\
  & &  \vec{R}_{-\vec{q},j} \, \cdotp \, \vec{P}_{vc,s}(\vec{k}_{\parallel}-\lambda_{h}\vec{q}_{\parallel}, \vec{k}_{\parallel} + \lambda_{e}\vec{q}_{\parallel} )  \psi_{\vec{q}_{\parallel},s,\alpha}(\vec{k}_{\parallel})  B_{\vec{q}_{\parallel},s,\alpha} \nonumber \\
  \label{eq:light_H2}
\end{eqnarray} 
and $ H^{+}_{j,s}=  [H^{-}_{j,s}]^{\dagger}$. We have expressed the field wavevector $\vec{q}$ in terms of the in-plane component, $\vec{q}_{\parallel}$, and the out-of-plane component, $q_{z}\hat{z}$.

\subsubsection{Transformation to Decoupled Longitudinal and Transverse Exciton Basis}
Excitons belong to a particular valley ($K$ or $K'$) are not the eigenstates in the presence of long-range dipole-dipole interactions~\cite{Wang16}. It is therefore convenient to switch excitons basis. Transverse and longitudinal exciton states, which are superspositions of exciton states belonging to the two valleys, are a much better choice. Transverse exciton exciton states couple with only TE polarized (or s-polarized) radiation and longitudinal exciton states couple with only TM polarized (or p-polarized) radiation~\cite{Wang16}. We define creation operators for transverse ($T$) and longitudinal ($L$) exciton states as follows,    
\begin{eqnarray}
 B^{\dagger}_{\vec{q}_{\parallel},L,\alpha} & = & \frac{1}{\sqrt{2}}  \left[ e^{-i\phi_{\vec{q}_{\parallel}}} B^{\dagger}_{\vec{q}_{\parallel},\tau=1,\alpha} + e^{i\phi_{\vec{q}_{\parallel}}} B^{\dagger}_{\vec{q}_{\parallel},\tau=-1,\alpha} \right] \nonumber \\
 B^{\dagger}_{\vec{q}_{\parallel},T,\alpha} &  = & \frac{i}{\sqrt{2}}  \left[ e^{-i\phi_{\vec{q}_{\parallel}}} B^{\dagger}_{\vec{q}_{\parallel},\tau=1,\alpha} - e^{i\phi_{\vec{q}_{\parallel}}} B^{\dagger}_{\vec{q}_{\parallel},\tau=-1,\alpha} \right] \nonumber \\
\end{eqnarray}
We also define exciton operator $C_{\vec{q}_{\parallel},L/T,\alpha}$ and photon operator $R_{\vec{q},L/T}$ as,
\begin{eqnarray}
  C_{\vec{q}_{\parallel},L/T,\alpha} & = & B_{\vec{q}_{\parallel},L/T,\alpha} + B^{\dagger}_{-\vec{q}_{\parallel},L/T,\alpha} \nonumber \\
  R_{\vec{q},L/T} & = & a_{\vec{q},L/T} +  a^{\dagger}_{-\vec{q},L/T} \label{eq:fields}
\end{eqnarray}
The optical matrix element can be expressed in terms of $\chi_{ex,\alpha}(\vec{r},\vec{q}_{\parallel})$,
\begin{eqnarray}
 \chi_{ex,\alpha}(\vec{r},\vec{q}_{\parallel}) & = &  \int \frac{d^{2}\vec{k}_{\parallel}}{(2\pi)^{2}} \vec{P}_{vc,s}(\vec{k}_{\parallel}-\lambda_{h}\vec{q}_{\parallel},\vec{k}_{\parallel}+\lambda_{e}\vec{q}_{\parallel}).\hat{x} \, \nonumber \\
   & & \times \psi_{\vec{q}_{\parallel},s,\alpha}(\vec{k}_{\parallel})  e^{i\vec{k}_{\parallel}.\vec{r}} \label{eq:chi}
\end{eqnarray}
Using these definitions, the exciton-photon Hamiltonian becomes,
\begin{equation}
H_{int} =  H_{L} +  H_{T} 
\end{equation}
where,
\begin{eqnarray}
 H_{L}  & = &  \frac{e}{m_{o}} \sum_{\vec{q},\alpha} \sqrt{\frac{A \hbar}{V \epsilon_{o} \omega_{q}}} e^{-iq_{z}z} \times \nonumber \\
  & & \chi_{ex,\alpha}(\vec{r}=0,\vec{q}_{\parallel}) \frac{|q_{z}|}{q} R_{-\vec{q},L} \, C_{\vec{q}_{\parallel},L,\alpha} \nonumber \\
 H_{T} &   = &  \frac{e}{m_{o}} \sum_{\vec{q},\alpha} \sqrt{\frac{A \hbar}{V \epsilon_{o} \omega_{q}}} e^{-iq_{z}z} \times \nonumber \\
 & & \chi_{ex,\alpha}(\vec{r}=0,\vec{q}_{\parallel}) R_{-\vec{q},T} \, C_{\vec{q}_{\parallel},T,\alpha} 
  \label{eq:light_H3}
\end{eqnarray}
The basis transformation decouples the longitudinal and transverse exciton-polaritons. In what follows we will assume that $\chi_{ex,\alpha}(0,\vec{q}_{\parallel})$ is real. If it had a phase, it could be absorbed in $\phi_{\vec{q}_{\parallel}}$ used to define the operators $B_{\vec{q}_{\parallel},L/T,\alpha}$ above.    

\subsubsection{The Quadratic Part of the Interaction Hamiltonian} \label{sec:quad}
The interaction Hamiltonian used above ignores the quadratic vector potential terms~\cite{Girlanda95,Wang16}. Following Girlanda et~al.~\cite{Girlanda95}, the form of these extra term is found to be,
\begin{eqnarray}
 & & \hat{H}'_{int} =  \hat{H}'_{L} + \hat{H}'_{T} \nonumber \\
  & & \hat{H}'_{L} = \sum_{\vec{Q},\alpha} E_{ex,\alpha} (\vec{Q}) \nonumber \\
  & & \times \left[ \frac{e}{m_{o}} \sum_{\vec{q}} \delta_{\vec{q}_{\parallel},\vec{Q}} \sqrt{\frac{A\hbar}{V \epsilon_{o} \omega_q}} e^{iq_{z}z} \frac{|\chi_{ex}(0,\vec{q}_{\parallel})|}{E_{ex,\alpha}(\vec{q}_{\parallel})} \frac{|q_{z}|}{q} R_{\vec{q},L}  \right] \nonumber \\
  & & \times \left[ \frac{e}{m_{o}} \sum_{\vec{k}} \delta_{\vec{k}_{\parallel},\vec{Q}} \sqrt{\frac{A\hbar}{V \epsilon_{o} \omega_k}} e^{-ik_{z}z} \frac{|\chi_{ex}(0,\vec{k}_{\parallel})|}{E_{ex,\alpha}(\vec{k}_{\parallel})} \frac{|k_{z}|}{k} R_{-\vec{k},L}  \right] \nonumber \\
   & & \hat{H}'_{T} = \sum_{\vec{Q},\alpha} E_{ex,\alpha} (\vec{Q}) \nonumber \\
  & & \times \left[ \frac{e}{m_{o}} \sum_{\vec{q}} \delta_{\vec{q}_{\parallel},\vec{Q}} \sqrt{\frac{A\hbar}{V \epsilon_{o} \omega_q}} e^{iq_{z}z} \frac{|\chi_{ex}(0,\vec{q}_{\parallel})|}{E_{ex,\alpha}(\vec{q}_{\parallel})}  R_{\vec{q},T}  \right] \nonumber \\
  & & \times \left[ \frac{e}{m_{o}} \sum_{\vec{k}} \delta_{\vec{k}_{\parallel},\vec{Q}} \sqrt{\frac{A\hbar}{V \epsilon_{o} \omega_k}} e^{-ik_{z}z} \frac{|\chi_{ex}(0,\vec{k}_{\parallel})|}{E_{ex,\alpha}(\vec{k}_{\parallel})}  R_{-\vec{k},T}  \right] \nonumber \\
  \end{eqnarray}
These terms must be added to the interaction Hamiltonian when expressions for the exciton self-energies are calculated~\cite{Wang16}.

\subsection{Exciton Spectral Density Functions}
We define the exciton Green's function $G^{<}_{\vec{Q},L/T,\alpha}(t-t')$ as follows~\cite{Mahan00},
\begin{equation}
G^{<}_{\vec{Q},L/T,\alpha}(t-t') = -\frac{i}{\hbar} \langle B^{\dagger}_{\vec{Q},L/T,\alpha}(t') B_{\vec{Q},L/T,\alpha}(t) \rangle
\end{equation}
The angular brackets indicate averaging with respect to an ensemble of excitons. In the frequency domain,
\begin{equation}
 G^{<}_{\vec{Q},L/T,\alpha}(\omega) = -\frac{i}{\hbar} A_{\vec{Q},L/T,\alpha}(\omega) n^{B}_{\vec{Q},L/T,\alpha}(\omega)   
\end{equation}
Here, $A_{\vec{Q},L/T,\alpha}(\omega)$ is the exciton spectral density function and $n^{B}_{\vec{Q},L/T,\alpha}(\omega)$ is the exciton occupation factor and equals the Bose-Einstein factor in thermal equilibrium. Most other exciton Green's functions can be obtained from the spectral density function~\cite{Mahan00}, which incorporates effects due to exciton-photon interaction as well as intra-layer exciton scattering. The average exciton number is,
\begin{eqnarray}
  \langle n_{\vec{Q},L/T,\alpha} \rangle & = & \langle B^{\dagger}_{\vec{Q},L/T,\alpha} B_{\vec{Q},L/T,\alpha} \rangle \nonumber \\
  & = & \int \frac{d\omega}{2\pi} \, \, i \hbar \, G^{<}_{\vec{Q},L/T,\alpha}(\omega)
\end{eqnarray}
The spectral density functions satisfy the sum rule,
\begin{equation}
\int \frac{d\omega}{2\pi} \, A_{\vec{Q},L/T,\alpha}(\omega) = 1
\end{equation}
The energy dispersions $E_{ex,L/T,\alpha}(\vec{Q})$ and the spectral density functions of the transverse and longitudinal excitons in the presence of exciton-photon interaction can be found from the corresponding retarded Green's functions~\cite{Wang16}, as shown in the Appendix. A convenient phenomenological choice for $ A_{\vec{Q},L/T,\alpha}(\omega)$ is a Lorentzian,
\begin{equation}
  A_{\vec{Q},L/T,\alpha}(\omega) = \frac{2\hbar\Gamma_{\vec{Q},L/T,\alpha}}{(\hbar \omega - E_{ex,L/T,\alpha}(\vec{Q}))^{2} + \Gamma^{2}_{\vec{Q},L/T,\alpha}}
\end{equation}
The FWHM exciton linewidth is $2\Gamma_{\vec{Q},L/T,\alpha}$. The one instance where the above simple Lorentzian form does not work well is in the case of the transverse excitons right when the spectral weight shifts between the two branches of the polariton dispersion when moving from inside the light cone to outside the light cone (see the Appendix).

\section{Rates of Exciton Energy Transfer} \label{sec:rate}
\begin{figure}
  \begin{center}
   \epsfig{file=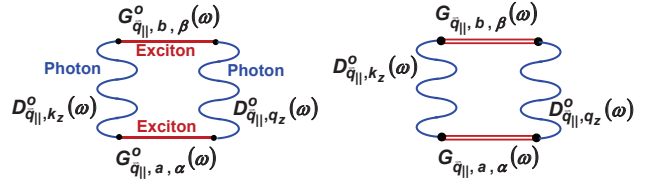,angle=0,width=0.48\textwidth}
   \caption{Feynman diagrams representing the transfer of energy between an exciton in layer $a$ and an exciton in layer $b$ by photon exchange. Green's functions of excitons (straight lines) and photons (wavy lines) are shown in the Figure. The subscripts $a$ and $b$ stand for layer $a$ and layer $b$, respectively. Internal variables, $q_{z}$ and $k_{z}$, are integrated over. (Left) Processes involving bare Green's functions. (Right) Dressed Green's functions.} 
    \label{fig:fig2}
  \end{center}
\end{figure}

We consider two parallel (not necessarily identical) electronically decoupled (but electromagnetically coupled) TMD monolayers, labeled $a$ and $b$, located at $z=0$ and $z=d$, respectively, as shown in the Figure~\ref{fig:fig1}. We assume that the exciton intralayer scattering and dephasing rates are fast so that energy transfer dynamics can be described as a simple decay. The case where exciton scattering and dephasing rates are slow and coherent dynamics are important is discussed later in this paper in Section~\ref{sec:coh}. We calculate the average of the rate of change of the number of excitons with in-plane momentum $\vec{Q}$ in layer $a$ as result of electromagnetic coupling to layer $b$. The desired Heisenberg operator is,
\begin{eqnarray}
  \dot{n}_{\vec{Q},L/T,a,\alpha} & = &  \frac{d \, B^{\dagger}_{\vec{Q},L/T,a,\alpha} B_{\vec{Q},L/T,a,\alpha}}{dt} \nonumber \\
 &  = & -\frac{i}{\hbar}[n_{\vec{Q},L/T,a,\alpha},H_{int} + H'_{int}]
\end{eqnarray}
Note that the layer index ($a$ or $b$) is added to the subscripts. We calculate the average of the operator $\dot{n}_{\vec{Q},L/T,a,\alpha}$ using the non-equilibrium Green's function technique~\cite{Mahan00},
\begin{eqnarray}
& &   \langle   \dot{n}_{\vec{Q},L/T,a,\alpha}(t) \rangle  =  \nonumber \\
  & & \langle T_{c} \left[ e^{-\frac{i}{\hbar}\int_{c} \left( H_{int}(t') + H'_{int}(t') \right) dt'} \dot{n}_{\vec{Q},L/T,a,\alpha}(t) \right] \rangle
\end{eqnarray}
Here, $T_{c}$ stands for operator contour ordering along the Keldysh contour $c$ that runs from time $-\infty$ to $+\infty$ and back~\cite{Mahan00}. The angled brackets stand for averaging with respect to the initial density matrix at time  $-\infty$~\cite{Mahan00}. The above expression can be evaluated in terms of Green's functions using standard perturbation techniques~\cite{Mahan00}. The lowest order non-zero terms in the perturbative expansion above give the rate of decrease of the exciton number due to spontaneous emission into free-space, as discussed by Wang et~al.~\cite{Wang16}. The terms relevant to the present discussion correspond to the Feynman diagrams depicted in Figure~\ref{fig:fig2} which represent energy transfer between the excitons in the two layers. The bare exciton Green's functions of each layer in Figure~\ref{fig:fig2} are dressed from, (i) intra-layer photon interactions (or intra-layer long-range dipole-dipole interactions), as discussed by Wang et~al.~\cite{Wang16}, and from (ii) intra-layer interactions responsible for exciton scattering assuming these interactions are included in the Hamiltonian.  
The final result for the energy transfer rate $R_{E}$ can be written in terms of the spectral density functions of the excitons in the two layers. For the transverse excitons we get,
\begin{eqnarray}
    & &   R_{E} = \langle   \dot{n}_{\vec{Q},T,a,\alpha}(t) \rangle  =  - \frac{1}{\hbar^{2}} \sum_{\beta} \int \frac{d\omega}{2\pi}   \nonumber \\
    & & \times A_{\vec{Q},T,a,\alpha}(\omega) A_{\vec{Q},T,b,\beta}(\omega)  \nonumber \\
    & & \times \left| \eta_{o} \frac{e^{2}}{m^{2}_{o}} \chi_{ex,a,\alpha}(0,\vec{Q}) \chi_{ex,b,\beta}(0,\vec{Q}) \right|^{2}   \frac{ \left| e^{i\sqrt{\omega^{2} - Q^{2}c^{2}}d/c} \right|^{2}}{| \omega^{2} - Q^{2}c^{2}| } \nonumber \\
    & & \times \left[  n^{B}_{\vec{Q},T,a,\alpha}(\omega) - n^{B}_{\vec{Q},T,b,\beta}(\omega) \right]  \label{eq:mainT}
\end{eqnarray}
and for the longitudinal excitons we obtain,
\begin{eqnarray}
    & &   R_{E} = \langle   \dot{n}_{\vec{Q},L,a,\alpha}(t) \rangle  =  - \frac{1}{\hbar^{2}} \sum_{\beta} \int \frac{d\omega}{2\pi}   \nonumber \\
    & & \times A_{\vec{Q},L,a,\alpha}(\omega) A_{\vec{Q},L,b,\beta}(\omega)  \nonumber \\
  & & \times \left| \eta_{o} \frac{e^{2}}{m^{2}_{o}} \chi_{ex,a,\alpha}(0,\vec{Q}) \chi_{ex,b,\beta}(0,\vec{Q}) \right|^{2}  \left| e^{i\sqrt{\omega^{2} - Q^{2}c^{2}}d/c} \right|^{2} \nonumber \\
  & & \times \frac{|\omega^{2} - Q^{2}c^{2}|}{\omega^{4}}  \left[  n^{B}_{\vec{Q},L,a,\alpha}(\omega) - n^{B}_{\vec{Q},L,b,\beta}(\omega) \right] \label{eq:mainL}
\end{eqnarray}
Note that the expressions above are valid for $\omega > Qc$ (radiative transfer) as well as for $\omega < Qc$ (non-radiative transfer) provided in the latter case the replacement $\sqrt{\omega^{2} - Q^{2}c^{2}} \rightarrow  i \sqrt{Q^{2}c^{2} - \omega^{2}}$ is made. The above expressions represent the main results of this work.   

\section{Discussion} \label{sec:dis}
The following points regarding the expressions above need to be noted,
\begin{enumerate}
\item The rate of energy transfer depends on the overlap of the spectral density functions of the excitons in the two layers.
\item Outside the light cone, when $\omega < Qc$, the energy transfer is mediated via evanescent fields and the rate of transfer decreases exponentially with interlayer separation $d$ as $\sim e^{-2\sqrt{Q^{2}c^{2}-\omega^{2}}d/c}$.
\item The energy transfer rates depend inversely on the exciton linewidth (and, therefore, the exciton intralayer scattering rates) via the exciton spectral density functions. 
\item The energy transfer rates depend on the dielectric constants (or the refractive indices) of the media surrounding the two monolayers. In the simple case when the media on either side of the layers and also in between the layers have the same dispersionless refractive index $n$, the impedance $\eta_{o}$ and the speed of light $c$ in the above expressions get replaced by $\eta_{o}/n$ and $c/n$, respectively.  
\item Since the exciton state belonging to one valley can be considered a superposition of transverse and longitudinal exciton states, its energy transfer rate will be the average of the energy transfer rates for the transverse and longitudinal excitons.   
\item In the static limit, $Qc >> \omega$, the expressions for the energy transfer rates obtained for the longitudinal excitons have the same form as those obtained previously for quantum well excitons using the static dipole-dipole interaction Hamiltonian~\cite{Lyo00}, which is to be expected.    
\end{enumerate}

\section{Numerical Results for the Exciton Energy Transfer Times Between Two MoS$_{2}$ Monolayers} \label{sec:num}
For numerical evaluations of the results, we assume two identical and parallel MoS$_{2}$ layers at a distance $d$. The electronic and optical parameter values used for MoS$_{2}$ monolayers are the same as those given previously~\cite{Wang16,Wang15b,Changjian14}. We first calculate the energy dispersions for the lowest energy $1s$ longitudinal and transverse excitons (as described in the Appendix), and then use these to compute the energy transfer rates. In numerical calculations, a momentum-independent scattering-limited FWHM exciton linewidth of $\sim$30 meV is assumed.   
\begin{figure}
  \begin{center}
   \epsfig{file=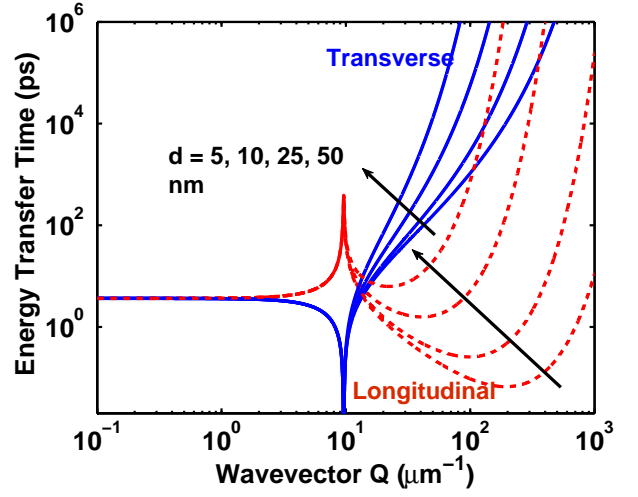,angle=0,width=0.48\textwidth}
   \caption{The calculated energy transfer times for the lowest energy $1s$ longitudinal (red dashed)  and transverse (blue solid) excitons in two parallel MoS$_{2}$ monolayers are plotted as a function of the exciton in-plane momentum $Q$ for different values of the interlayer separation $d$ ($d$=5, 10, 25 and 50 nm). The value of the momentum $Q_{o}$, defined by $\hbar Q_{o} c = E_{ex,1s}(Q_{o})$, is $\sim$9.6 1/$\mu$m (see the Appendix). A FWHM exciton linewidth of $\sim$30 meV is assumed.} 
    \label{fig:fig3}
  \end{center}
\end{figure}
Figure~\ref{fig:fig3} shows the calculated energy transfer times for both longitudinal and transverse excitons as a function of the exciton in-plane momentum $Q$ for different values of the interlayer separation $d$ ($d$=5, 10, 25 and 50 nm). The value of the momentum $Q_{o}$, defined by $\hbar Q_{o} c = E_{ex,1s}(Q_{o})$, is $\sim$9.6 1/$\mu$m (see the Appendix). For $Q<<Q_{o}$ (inside the light cone), the energy transfer is via the radiative mechanism and the energy transfer time is almost independent of $Q$. When $Q \sim Q_{o}$, the cusps in the energy transfer times follow the trends in the radiative lifetimes and the optical conductivities of longitudinal and transverse excitons (see the Appendix). When $Q>Q_{o}$ (outside the light cone), the energy transfer is via the non-radiative mechanism (via evanescent waves). In the case of the transverse excitons, the energy transfer rate decreases (and the energy transfer time increases) exponentially with the product $\sim Qd$ (for large $Q$). In the case of the longitudinal excitons, the energy transfer time first decreases with $Q$ because the in-plane component of the evanescent radiation also increases with $Q$, and then the energy transfer time increases exponentially with the product $\sim Qd$ (for large $Q$). Note that for interlayer separations $d$ less than 10 nm, the energy transfer times for the longitudinal excitons can be in the hundreds of femtoseconds range or even smaller. These results show the efficacy of the exciton energy transfer mechanism in TMD monolayers.    

Simple expressions for the energy transfer times $\tau_{E,\vec{Q},L/T,\alpha}$ can be found for two identical TMD layers when $Q<Q_{o}$ and $Q>Q_{o}$ (i.e. away from $Q=Q_{o}$) assuming Lorentzian spectral density functions,
\begin{eqnarray}
\frac{1}{\tau_{E,\vec{Q},T,\alpha}} & \approx &   \frac{\hbar}{2 \gamma} \left(2\eta_{o} \frac{e^{2}}{m^{2}_{o}} |\chi_{ex,\alpha}(0,\vec{Q})|^{2} \right)^{2} \nonumber \\
& & \times \frac{ e^{2i\sqrt{f_{\vec{Q},T,\alpha}} d/\hbar c}}{|f_{\vec{Q},T,\alpha}|} \label{eq:simple1} \\
\frac{1}{\tau_{E,\vec{Q},L,\alpha}} & \approx &   \frac{\hbar}{2 \gamma} \left(2\eta_{o} \frac{e^{2}}{m^{2}_{o}} |\chi_{ex,\alpha}(0,\vec{Q})|^{2} \right)^{2} \nonumber \\
& & \times e^{2i\sqrt{f_{\vec{Q},L,\alpha}} d/\hbar c} \frac{|f_{\vec{Q},L,\alpha}|}{E^{4}_{ex,L,\alpha}(\vec{Q})} \label{eq:simple2}
  \end{eqnarray}
where $\gamma = (\Gamma_{\vec{Q},L/T,a,\alpha} + \Gamma_{\vec{Q},L/T,b,\alpha})$, $f_{\vec{Q},L/T,\alpha} = (E^{2}_{ex,L/T,\alpha}(\vec{Q}) - \hbar^{2}Q^{2}c^{2})$, and $\sqrt{f_{\vec{Q},L/T,\alpha}} = i\sqrt{|f_{\vec{Q},L/T,\alpha}|}$ when $f_{\vec{Q},L/T,\alpha}<0$ outside the light cone. For  $Q<Q_{o}$ (radiative energy transfer) the above expressions can be written as,
\begin{equation}
  \frac{1}{\tau_{E,\vec{Q},L/T,\alpha}} \approx \frac{\hbar}{2 \gamma} \left( \frac{1}{\tau_{sp,\vec{Q},L/T,\alpha}} \right)^{2} \label{eq:simple0}
\end{equation}
Here, $\tau_{sp,\vec{Q},L/T,\alpha}$ is the spontaneous emission radiative lifetime of excitons in a single TMD layer~\cite{Wang16} (see the Appendix). In the case of two MoS$_{2}$ monolayers, $\tau_{sp,\vec{Q}\approx 0,L/T,1s}$ is around 200 fs (for $1s$ excitons)~\cite{Wang16}. Assuming FWHM exciton linewidths of $\sim$30 meV~\cite{Wang15c,Changjian14} and $\sim$10 meV, the radiative energy transfer times are $\sim$3.6 ps and $\sim$1.2 ps, respectively. Although these times seem short, the radiative energy transfer process will not be very efficient (efficiency less than $\sim$15\%) because the spontaneous emission time is also very short and only a small fraction of the photons emitted from one layer get absorbed by the other layer.

\begin{figure}
  \begin{center}
   \epsfig{file=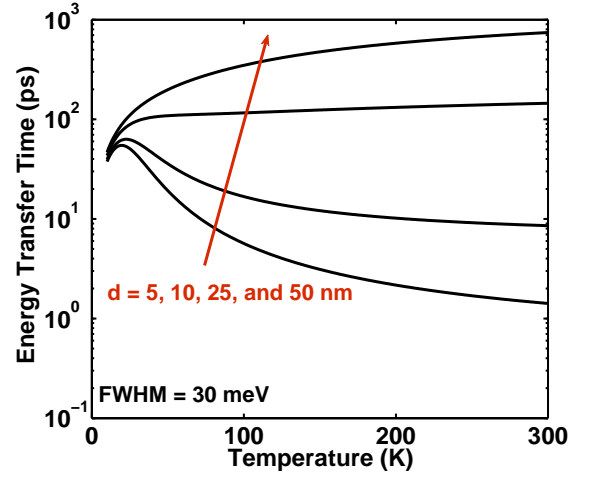,angle=0,width=0.44\textwidth}
   \caption{The calculated average energy transfer times for a thermal ensemble of ($1s$) longitudinal and transverse excitons in two parallel MoS$_{2}$ monolayers are plotted as a function of the exciton temperature for different for different values of the interlayer separation $d$ ($d$=5, 10, 25 and 50 nm). The exciton FWHM linewidth is assumed to be 30 meV.} 
    \label{fig:fig4a}
  \end{center}
\end{figure}

\begin{figure}
  \begin{center}
   \epsfig{file=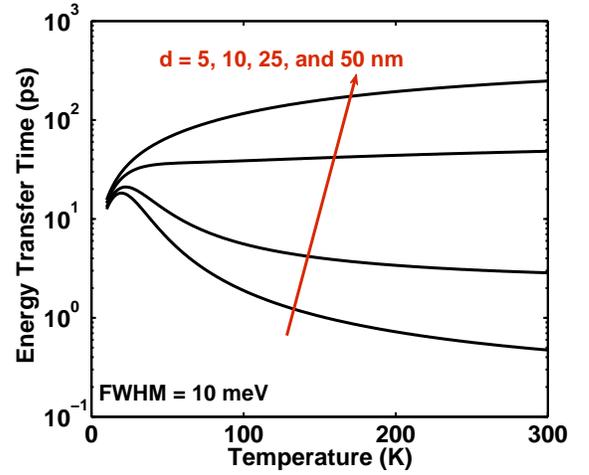,angle=0,width=0.44\textwidth}
   \caption{The calculated average energy transfer times for a thermal ensemble of ($1s$) longitudinal and transverse excitons in two parallel MoS$_{2}$ monolayers are plotted as a function of the exciton temperature for different for different values of the interlayer separation $d$ ($d$=5, 10, 25 and 50 nm). The exciton FWHM linewidth is assumed to be 10 meV.} 
    \label{fig:fig4b}
  \end{center}
\end{figure}

Despite the fast energy transfer rates for the longitudinal excitons when $Q > Q_{o}$, their contribution to the energy transfer process is expected to be limited by the relatively small density of the longitudinal excitons in a thermal ensemble of excitons since $E_{ex,L,1s}(\vec{Q}) > E_{ex,T,1s}(\vec{Q})$ for $Q > Q_{o}$ (see the Appendix). Figure~\ref{fig:fig4a} shows the calculated average energy transfer times for a thermal ensemble of ($1s$) longitudinal and transverse excitons as a function of the exciton temperature. The exciton density is assumed to be dilute enough such that the exciton chemical potential is less than the lowest exciton energy level by at least several $KT$. The exciton FWHM linewidth is assumed to be momentum-independent and equal to 30 meV. Figure~\ref{fig:fig4b} shows the same results assuming that the exciton FWHM linewidth is 10 meV. The results show that when the interlayer separation is small then as the temperature increases, and the density of the longitudinal excitons also increases relative to the transverse excitons, the average energy transfer time decreases. However, when the interlayer separation is large, and the energy transfer is by excitons with only small momenta ($Q < 1/d$), an increase of the temperature results in an increase of the energy transfer time because the exciton thermal distribution spills to larger momenta. These results also show that the average energy transfer times scale with the exciton FWHM linewidth and can be shorter than a picosecond for interlayer separations smaller than 10 nm and exciton linewidths narrower than 10 meV.

\section{Coherent Energy Transfer Dynamics} \label{sec:coh}
In the previous section we assumed that the intralayer exciton scattering and dephasing rates are slow and energy transfer dynamics can be described as a simple decay. Here we quantify this notion and also discuss coherent interlayer energy transfer dynamics. First, we evaluate corrections to the exciton dispersions as a result of interlayer radiative and non-radiative interactions. 

We consider two parallel and {\em identical} electronically decoupled (but electromagnetically coupled) TMD monolayers, labeled $a$ and $b$, located at $z=0$ and $z=d$, respectively, as shown earlier in Figure~\ref{fig:fig1}. The analysis is greatly simplified if we define operators for the in-phase ('+' exciton) and out-of-phase ('-' exciton) excitons in the two layers as follows~\cite{Citrin94},
\begin{equation}
  B_{\vec{q}_{\parallel},L/T,\pm,\alpha} = \frac{B_{\vec{q}_{\parallel},L/T,a,\alpha} \pm B_{\vec{q}_{\parallel},L/T,b,\alpha}}{\sqrt{2}}
\end{equation}
The '+' and '-' excitons have their dipole moments in-phase and out-of-phase, respectively. The retarded Green's functions and self-energies for the '+' and '-' excitons can be found using the methods described in the Appendix,
\begin{eqnarray}
  & & G^{R}_{\vec{q}_{\parallel},L/T,\pm,\alpha}(\omega) = \nonumber \\
  & = & \frac{2E_{ex,\alpha}(\vec{q}_{\parallel})  \left[ 1-  2 \, \Sigma^{oR}_{\vec{q}_{\parallel},L/T,\pm,\alpha}(\omega)/E_{ex,\alpha}(\vec{q}_{\parallel}) \right]        }{(\hbar\omega)^{2}-E_{ex,\alpha}(\vec{q}_{\parallel})^{2} - 2\frac{(\hbar \omega)^{2}}{E_{ex,\alpha}(\vec{q}_{\parallel})} \Sigma^{oR}_{\vec{q}_{\parallel},L/T,\pm,\alpha}(\omega)} \nonumber \\
  & \approx & \frac{2E_{ex,\alpha}(\vec{q}_{\parallel})}{(\hbar\omega)^{2}-E_{ex,\alpha}(\vec{q}_{\parallel})^{2} - 2\frac{(\hbar \omega)^{2}}{E_{ex,\alpha}(\vec{q}_{\parallel})} \Sigma^{oR}_{\vec{q}_{\parallel},L/T,\pm,\alpha}(\omega)} \nonumber \\
\end{eqnarray}
Here,
\begin{equation}
  \Sigma^{oR}_{\vec{q}_{\parallel},L/T,\pm,\alpha}(\omega) = \Sigma^{oR}_{\vec{q}_{\parallel},L/T,\alpha}(\omega) \left[ 1 \pm e^{i\sqrt{\omega^{2} - q^{2}_{\parallel}c^{2}}d/c} \right]
\end{equation}
$\Sigma^{oR}_{\vec{q}_{\parallel},L/T,\alpha}(\omega)$ is the retarded self-energy for excitons in a single TMD layer and its expression was given previously~\cite{Wang16} (also see the Appendix). The expression above is valid for $\omega > q_{\parallel}c$ as well as for $\omega <  q_{\parallel}c$ provided in the latter case the replacement $\sqrt{\omega^{2} -  q_{\parallel}^{2}c^{2}} \rightarrow  i \sqrt{ q_{\parallel}^{2}c^{2} - \omega^{2}}$ is made. It is clear from the above expression for the self-energy that in the limit $d \rightarrow 0$ the out-of-phase '-' excitons do not radiate whereas the radiative rates of the in-phase '+' excitons are twice as fast as those of excitons in a single TMD layer. The energy splitting between  the '+' and '-' excitons due to interlayer interactions can be estimated as,
\begin{eqnarray}
& &   \Delta_{\vec{q}_{\parallel},L/T,\alpha} =  \nonumber \\
        & & {\rm Real} \left[ 2\Sigma^{oR}_{\vec{q}_{\parallel},L/T,\alpha}(\omega) e^{i\sqrt{\omega^{2} - q^{2}_{\parallel}c^{2}}d/c} \, \right]_{\hbar \omega = E_{ex,L/T,\alpha}(\vec{q}_{\parallel})} \nonumber \\
\end{eqnarray}
Since an exciton state in any one of the two TMD layers can be considered a superposition of the in-phase and the out-of-phase exciton states, if the energy splitting $\Delta_{\vec{q}_{\parallel},L/T,\alpha}$ due to interlayer interactions is much larger than the exciton linewidth due to intralayer scattering and dephasing then coherent energy oscillations between exciton states in the two layers are expected at the frequency $\Delta_{\vec{q}_{\parallel},L/T,\alpha}/\hbar$, and energy transfer between the layers cannot be described as a simple decay of energy from one layer to the other.

\begin{figure}
  \begin{center}
   \epsfig{file=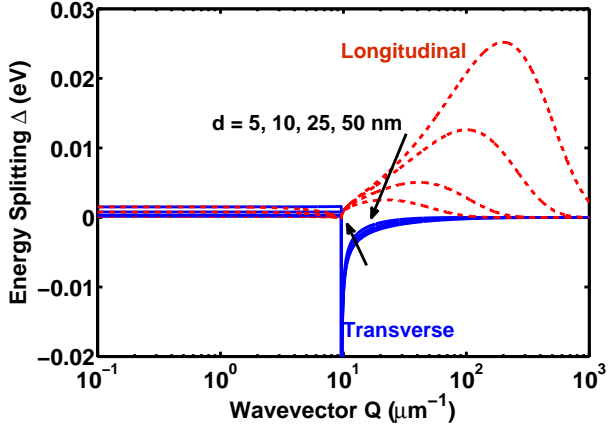,angle=0,width=0.48\textwidth}
   \caption{Calculated energy splittings between the '+' and '-' excitons are plotted for the transverse (blue-solid) and longitudinal (red-dashed) excitons for the lowest energy ($1s$) exciton state in two suspended MoS$_{2}$ monolayers as a function of the in-plane momentum $Q$ for different values of the interlayer spacing $d$. The value of the momentum $Q_{o}$, defined by $\hbar Q_{o} c = E_{ex,1s}(Q_{o})$, is $\sim$9.6 1/$\mu$m }
    \label{fig:FigA2}
  \end{center}
\end{figure}

The calculated energy splittings between the '+' and '-' excitons are plotted in Figure~\ref{fig:FigA2} for the lowest energy ($1s$) exciton state in two suspended MoS$_{2}$ monolayers as a function of the in-plane momentum $Q$ for different values of the interlayer spacing $d$ ($d$=5, 10, 25 and 50 nm). Inside the light cone, the energy splittings are small (less than 1-2 meV) for all values of $d$ considered. The energy splittings are smaller than the natural linewidth of excitons in a single MoS$_{2}$ layer due to radiative decay. Consequently, coherent energy oscillations are not expected for excitons inside the light cone. Outside the light cone, the energy splittings for the longitudinal excitons become large reaching values larger than $\sim$25 meV for $d$ less than 5 nm. However, these large energy splittings occur at large values of the exciton momenta where exciton intralayer scattering is also expected to be fast and the condition for coherent oscillations might be difficult to meet. If, outside the light cone, exciton scattering rates are small then the coherent dynamics of the average exciton layer number $\langle n_{\vec{Q},L/T,a/b,\alpha} \rangle$ can be modeled by the following equation similar to that of a damped simple harmonic oscillator,
\begin{eqnarray}
  & & \left[ \frac{d^{2}}{dt^{2}} + \frac{\gamma}{\hbar}\frac{d}{dt} + \frac{\Delta^{2}_{\vec{Q},L/T,\alpha}}{\hbar^{2}} \right] \langle n_{\vec{Q},L/T,a/b,\alpha}(t) \rangle = \nonumber \\
  & & \frac{1}{2} \frac{\Delta^{2}_{\vec{Q},L/T,\alpha}}{\hbar^{2}}
\end{eqnarray}
Here, $\gamma = (\Gamma_{\vec{Q},L/T,a,\alpha} + \Gamma_{\vec{Q},L/T,b,\alpha})$ is related to the exciton scattering and dephasing rate and the appropriate boundary conditions are,
\begin{eqnarray}
& & n_{\vec{Q},L/T,a,\alpha}(t=0) = 1 \nonumber \\
& & n_{\vec{Q},L/T,b,\alpha}(t=0) = 0 
\end{eqnarray}
If $\Delta_{\vec{Q},L/T,\alpha} > \gamma/2$, then the above equation predicts damped oscillations at the frequency $\Delta_{\vec{Q},L/T,\alpha}/\hbar$. On the other hand, if $\Delta_{\vec{Q},L/T,\alpha} << \gamma/2$, then the above equation gives a simple exponential decay of energy from layer $a$ to layer $b$ at the rate $\Delta^{2}_{\vec{Q},L/T,\alpha}/(2\hbar \gamma)$. This latter result is almost exactly what was obtained earlier in (\ref{eq:simple1}) and (\ref{eq:simple2}). In the absence of quantitative models or experimental data for exciton scattering in TMDs it is difficult to say if coherent energy oscillations are possible in TMDs. In the case of localized excitons, discussed next, momentum spread due to localization also contributes to the decoherence of the oscillations.

\section{Energy Transfer Rates for Localized Excitons}  \label{sec:app1}

\subsection{Energy Transfer Rates for a Localized Exciton in Layer $a$ and Free Excitons in Layer $b$}
The analysis in the previous Sections shows that the longitudinal excitons with momenta $Q$ in the $Q_{o} < Q < 1/d$ range have the shortest energy transfer times but the density of such excitons is relatively small in a thermal ensemble at low temperatures thus limiting the average energy transfer rates. Localized excitons, whose wavefunction is a superposition of exciton states of different momenta, could overcome these limitations and exhibit fast energy transfer rates even at low temperatures. We consider an initial exciton state in layer $a$ that is localized in space in a region of size $L_{c}$. We assume that the exciton wavefunction for the center of mass coordinate in real and Fourier spaces is,
\begin{eqnarray} 
  & & \psi_{com}(\vec{R}) = \frac{1}{\sqrt{\pi L_{c}^{2}}} e^{-R^{2}/2L_{c}^{2}} \nonumber \\
  & & \psi_{com}(\vec{Q}) = \sqrt{4\pi L_{c}^{2}} e^{-Q^{2}L_{c}^{2}/2} \label{eq:psicom}
\end{eqnarray}
A localized exciton state can be constructed from the ground state $|\psi_{o} \rangle $, corresponding to a filled valence band and an empty conduction band, as follows~\cite{Wang16},
\begin{eqnarray}
|\psi_{L/T,a,\alpha} \rangle_{ex} & & =  \frac{1}{\sqrt{A}} \sum_{\vec{Q}} \psi_{com}(\vec{Q}) B^{\dagger}_{\vec{Q},L/T,a,\alpha} |\psi_{o} \rangle  \label{eq:exciton2}
\end{eqnarray}
Assuming the above localized state as the initial state, with a spectral density function $A_{L/T,a,\alpha}(\omega)$ with HWHM linewidth of $\Gamma_{L/T,a,\alpha}$ and centered at the energy $E_{ex,L/T,a,\alpha}$, the rate of energy transfer to free excitons in layer $b$ for the transverse case is found to be,
\begin{eqnarray}
    & &  R_{E}  =  - \frac{1}{\hbar^{2}} \sum_{\beta} \int \frac{d^{2}\vec{Q}}{(2\pi)^{2}} \, |\psi_{com}(\vec{Q})|^{2} \nonumber \\
& & \times  \, \int \frac{d\omega}{2\pi}  \, A_{T,a,\alpha}(\omega) A_{\vec{Q},T,b,\beta}(\omega)  \nonumber \\
    & & \times \left| \eta_{o} \frac{e^{2}}{m^{2}_{o}} \chi_{ex,a,\alpha}(0,\vec{Q}) \chi_{ex,b,\beta}(0,\vec{Q}) \right|^{2}   \frac{ \left| e^{i\sqrt{\omega^{2} - Q^{2}c^{2}}d/c} \right|^{2}}{| \omega^{2} - Q^{2}c^{2}| } \nonumber \\
    & & \times \left[ n^{B}_{T,a,\alpha}(\omega) - n^{B}_{\vec{Q},T,b,\beta}(\omega) \right] 
\end{eqnarray}
and for the longitudinal case we obtain,
\begin{eqnarray}
    & &  R_{E}  =  - \frac{1}{\hbar^{2}} \sum_{\beta} \int \frac{d^{2}\vec{Q}}{(2\pi)^{2}} \, |\psi_{com}(\vec{Q})|^{2} \nonumber \\
& & \times  \, \int \frac{d\omega}{2\pi}  \, A_{L,a,\alpha}(\omega) A_{\vec{Q},L,b,\beta}(\omega)  \nonumber \\
    & & \times \left| \eta_{o} \frac{e^{2}}{m^{2}_{o}} \chi_{ex,a,\alpha}(0,\vec{Q}) \chi_{ex,b,\beta}(0,\vec{Q}) \right|^{2}   \left| e^{i\sqrt{\omega^{2} - Q^{2}c^{2}}d/c} \right|^{2}  \nonumber \\
    & & \times \frac{|\omega^{2} - Q^{2}c^{2}|}{\omega^{4}}  \left[  n^{B}_{L,a,\alpha}(\omega) - n^{B}_{\vec{Q},L,b,\beta}(\omega) \right] 
\end{eqnarray}
Again note that the expressions above are valid for $\omega < Qc$ (non-radiative transfer) provided the replacement $\sqrt{\omega^{2} - Q^{2}c^{2}} \rightarrow  i \sqrt{Q^{2}c^{2} - \omega^{2}}$ is made.

Simpler expressions can be obtained in some special cases. Suppose there exists an exciton state $\beta$ with momentum $Q^{*}$ and a corresponding energy $E^{*}$ in layer $b$ which satisfies the energy conservation relation $E_{ex,L/T,b,\beta}(Q^{*}) = E_{ex,L/T,b,\beta}(Q=0) + E^{*} =  E_{ex.L/T,a,\alpha}$. If the exciton in layer $a$ is strongly localized such that the energy spread of the free excitons in layer $b$ corresponding to the momentum spread of the localized exciton in layer $a$ is much greater than $\gamma = (\Gamma_{L/T,a,\alpha} + \Gamma_{\vec{Q},L/T,b,\alpha})$, coherent oscillations will not be possible. If $Q^{*}<Q_{o}$ and $Q^{*}>Q_{o}$ (i.e. away from $Q^{*}=Q_{o}$) then, assuming Lorentzian spectral density functions, the energy transfer times can be written as,
\begin{eqnarray}
  \frac{1}{\tau_{E,T,\alpha}} & \approx &   \frac{\pi \hbar}{2} g_{ex,T,b,\beta}(E^{*}) |\psi_{com}(Q^{*})|^{2} \nonumber \\
  & & \times \left(2\eta_{o} \frac{e^{2}}{m^{2}_{o}} |\chi_{ex,a,\alpha}(0,\vec{Q}^{*})\chi_{ex,b,\beta}(0,\vec{Q}^{*})| \right)^{2} \nonumber \\
  & & \times \frac{ e^{2i\sqrt{f_{T,a,\alpha}} d/\hbar c}}{|f_{T,a,\alpha}|} \label{eq:simple3} \\
  \frac{1}{\tau_{E,L,\alpha}} & \approx &   \frac{\pi \hbar}{2} g_{ex,L,b,\beta}(E^{*}) |\psi_{com}(Q^{*})|^{2} \nonumber \\
& & \times \left(2\eta_{o} \frac{e^{2}}{m^{2}_{o}} |\chi_{ex,a,\alpha}(0,\vec{Q}^{*})\chi_{ex,b,\beta}(0,\vec{Q}^{*})| \right)^{2} \nonumber \\
& & \times e^{2i\sqrt{f_{L,a,\alpha}} d/\hbar c} \frac{|f_{L,a,\alpha}|}{E^{4}_{ex,L,a,\alpha}} \label{eq:simple4}
  \end{eqnarray}
where $f_{L/T,a,\alpha} = (E^{2}_{ex,L/T,a,\alpha} - \hbar^{2}Q^{2}c^{2})$, and $\sqrt{f_{L/T,a,\alpha}} = i\sqrt{|f_{L/T,a,\alpha}|}$ when $f_{L/T,a,\alpha}<0$ outside the light cone. Here, $g_{ex,L/T,b,\beta}(E)$ is the density of states of free excitons in layer $b$. Note that in this case the energy transfer times do not depend on the scattering/dephasing rate given by $\gamma$. Assuming $\alpha = \beta$, the expressions above differ from the corresponding expressions for free excitons, given earlier in (\ref{eq:simple1}) and (\ref{eq:simple2}), by a multiplicative factor of  $\pi \gamma g_{ex,L/T,b,\alpha}(E^{*}) |\psi_{com}(Q^{*})|^{2}$. For strongly localized excitons, this factor is of the order of unity and therefore energy transfer times for strongly localized $1s$ excitons in MoS$_{2}$, when plotted as a function of $Q^{*}$, are expected to be similar to those appearing in Figure (\ref{fig:fig3}). 

\subsection{Energy Transfer Rates for a Localized Exciton in Layer $a$ and a Localized Exciton in Layer $b$}
We now consider the case in which the final exciton state in layer $b$ is also localized. The center of mass wavefunctions of the excitons in layer $a$ and $b$ in real space are centered at the in-plane vectors $\vec{\rho}_{a}$ and $\vec{\rho}_{b}$, respectively, and in momentum space these wavefunctions are $\psi_{com,a}(\vec{q}_{\parallel})e^{-i\vec{q}_{\parallel}.\vec{\rho}_{a}}$ and $\psi_{com,b}(\vec{q}_{\parallel})e^{-i\vec{q}_{\parallel}.\vec{\rho}_{b}}$, respectively, where $\psi_{com,a/b}(\vec{q}_{\parallel})$ are as given earlier in (\ref{eq:psicom}). The vector $\vec{r}$ connects the center of the exciton states, $\vec{r} = (\vec{\rho}_{b} - \vec{\rho}_{a}) + d\hat{z}$. The rate of energy transfer for the transverse case is found to be,
\begin{eqnarray}
    & &  R_{E}  =  - \frac{1}{\hbar^{2}} \sum_{\beta} \int \frac{d\omega}{2\pi}  A_{T,a,\alpha}(\omega) A_{T,b,\beta}(\omega)  \nonumber \\
& & \times \left| \int \frac{d^{3}\vec{q}}{(2\pi)^{3}} \, \psi^{*}_{com,b}(\vec{q}_{\parallel})\psi_{com,a}(\vec{q}_{\parallel}) \, \, e^{i\vec{q}.\vec{r}} \, \frac{2c}{\omega^{2} - \omega^{2}_{q} + i\eta} \right. \nonumber \\
& & \times \left. \eta_{o} \frac{e^{2}}{m^{2}_{o}} \chi_{ex,a,\alpha}(0,\vec{q}_{\parallel}) \chi_{ex,b,\beta}(0,\vec{q}_{\parallel}) \right|^{2} \nonumber \\
    & & \times \left[  n^{B}_{T,a,\alpha}(\omega) - n^{B}_{T,b,\beta}(\omega) \right] 
\end{eqnarray}
and for the longitudinal case we get,
\begin{eqnarray}
    & &  R_{E}  =  - \frac{1}{\hbar^{2}} \sum_{\beta} \int \frac{d\omega}{2\pi}   A_{L,a,\alpha}(\omega) A_{L,b,\beta}(\omega)  \nonumber \\
& & \times \left| \int \frac{d^{3}\vec{q}}{(2\pi)^{3}} \, \psi^{*}_{com,b}(\vec{q}_{\parallel})\psi_{com,a}(\vec{q}_{\parallel}) \, \, e^{i\vec{q}.\vec{r}} \, \frac{2c}{\omega^{2} - \omega^{2}_{q} + i\eta}  \right. \nonumber \\
& & \times \left. \left( 1 - \frac{q^{2}_{\parallel}}{\omega^{2}/c^{2}} \right) \, \eta_{o} \frac{e^{2}}{m^{2}_{o}} \chi_{ex,a,\alpha}(0,\vec{q}_{\parallel}) \chi_{ex,b,\beta}(0,\vec{q}_{\parallel}) \right|^{2} \nonumber \\
    & & \times \left[ n^{B}_{L,a,\alpha}(\omega) - n^{B}_{L,b,\beta}(\omega) \right] 
\end{eqnarray}
An interesting case is that of extremely localized excitons for which $L_{c} << \hbar c/ E_{ex,L/T,a/b}$ and $L_{c} << d$. Assuming wavevector independent values of $\chi_{ex,a/b,\alpha/\beta}(0,\vec{q}_{\parallel})$ and using the results,
\begin{eqnarray}
& &  \int \frac{d^{3}\vec{q}}{(2\pi)^{3}} \, e^{i\vec{q}.\vec{r}} \, \frac{1}{\omega^{2} - \omega^{2}_{q} + i\eta} = -\frac{e^{i\frac{\omega}{c}r}}{4 \pi r c^{2}} \nonumber \\
& &  \int \frac{d^{3}\vec{q}}{(2\pi)^{3}} \, e^{i\vec{q}.\vec{r}} \, \frac{\left( 1 - \frac{q^{2}_{\parallel}}{\omega^{2}/c^{2}} \right)}{\omega^{2} - \omega^{2}_{q} + i\eta} = -\frac{e^{i\frac{\omega}{c}r}}{4 \pi r^{3} \omega^{2}} \left( 1 - i\frac{\omega}{c}r \right) \nonumber \\
\end{eqnarray}
the above expressions for $R_{E}$, in the limit $(\omega/c)r < < 1$, give a $1/r^{2}$ dependence of the energy transfer rate for the transverse case and a $1/r^{6}$ dependence for the longitudinal case. It is satisfying to note that the former result corresponds to the classical inverse square law for radiative energy transfer and the latter corresponds to the standard Forster's result for non-radiative energy transfer via dipole-dipole interaction~\cite{Andrews04,Forster48}. In the longitudinal case, if one integrates the energy transfer rate over the in-plane position $\vec{\rho}_{b}$ of the final exciton state, then the total energy transfer rate will scale as $1/d^{4}$ with the interlayer separation.

\section{Energy Transfer Rates for an Exciton in Layer $a$ and Free Electron-Hole Pairs in Layer $b$} \label{sec:app2}
In many cases of practical interest where the optical bandgaps and/or the exciton binding energies in two different TMD monolayers are very different, the energy transfer can be from the excitons in the wider bandgap TMD layer to the free electron-hole pairs in the narrower bandgap TMD layer. For example, this could be the case in two parallel monolayers of MoS$_{2}$ and MoTe$_{2}$~\cite{Yang15}.   

We assume that an exciton in layer $a$ with momentum $\vec{Q}$ decays into a free electron-hole pair in layer $b$. We assume that the energy emission and absorption remains close to the conduction and valence band edges at $K$ and $K'$ valleys so that the standard optical selection rules are not violated. Since a free electron-hole pair can be considered an unbound exciton, the expressions for the rate of energy transfer between excitons given in the main text are also valid for energy transfer between excitons in layer $a$ and free electron-hole pairs in layer $b$ provided the relative exciton wavefunction in layer $b$ is assumed to be a plane wave, the exciton energy dispersion in layer $b$ is replaced by that of a free electron-hole pair, and the summation over the final exciton states (i.e. over $\beta$) in layer $b$ is replaced by a phase space integral over the relative wavevector. 

We assume that the free electron-hole pair in layer $b$ is described by the spectral density function $A_{\vec{k},\vec{Q},b}(\omega)$ with HWHM linewidth $\Gamma_{\vec{k},\vec{Q},b}$. Here, $\vec{k}$ is the relative momentum of the electron-hole pair and $\vec{Q}$ is the center of mass momentum and the spectral density function is centered at the energy $E_{g,b} + \hbar^{2}k^{2}/2m_{r,b} + \hbar^{2}Q^{2}/2m_{ex,b}$. $m_{r,b}$ is the reduced electron-hole mass in layer $b$, $m_{ex,b}$ is the exciton mass in layer $b$, and $E_{g,b}$ is the bandgap of layer $b$.   

For the case of the transverse excitons in layer $a$ we get,
\begin{eqnarray}
    & &   R_{E} = \langle   \dot{n}_{\vec{Q},T,a,\alpha}(t) \rangle  =  - \frac{1}{\hbar^{2}} \int \frac{d^{2}\vec{k}}{(2\pi)^{2}} \int \frac{d\omega}{2\pi}   \nonumber \\
    & & \times A_{\vec{Q},T,a,\alpha}(\omega) A_{\vec{k},\vec{Q},b}(\omega)  \nonumber \\
    & & \times \left| \eta_{o} \frac{e^{2}}{m^{2}_{o}} \chi_{ex,a,\alpha}(0,\vec{Q}) \chi_{\vec{k},b}(\vec{Q}) \right|^{2}   \frac{ \left| e^{i\sqrt{\omega^{2} - Q^{2}c^{2}}d/c} \right|^{2}}{| \omega^{2} - Q^{2}c^{2}| } \nonumber \\
    & & \times \left\{  n^{B}_{\vec{Q},T,a,\alpha}(\omega) \left[ f_{v}(\vec{k} - \lambda_{h,b}\vec{Q}) - f_{c}(\vec{k} + \lambda_{e,b}\vec{Q}) \right] \right. \nonumber \\
& & \left. -  f_{c}(\vec{k} + \lambda_{e,b}\vec{Q}) \left[ 1 -  f_{v}(\vec{k} - \lambda_{h,b}\vec{Q}) \right] \right\} 
\end{eqnarray}
and for the longitudinal excitons we obtain,
\begin{eqnarray}
    & &   R_{E} = \langle   \dot{n}_{\vec{Q},L,a,\alpha}(t) \rangle  =  - \frac{1}{\hbar^{2}} \int \frac{d^{2}\vec{k}}{(2\pi)^{2}} \int \frac{d\omega}{2\pi}   \nonumber \\
    & & \times A_{\vec{Q},L,a,\alpha}(\omega)    A_{\vec{k},\vec{Q},b}(\omega) \nonumber \\
  & & \times \left| \eta_{o} \frac{e^{2}}{m^{2}_{o}} \chi_{ex,a,\alpha}(0,\vec{Q}) \chi_{\vec{k},b}(\vec{Q}) \right|^{2}  \left| e^{i\sqrt{\omega^{2} - Q^{2}c^{2}}d/c} \right|^{2} \nonumber \\
  & & \times \frac{|\omega^{2} - Q^{2}c^{2}|}{\omega^{4}}  \left\{  n^{B}_{\vec{Q},L,a,\alpha}(\omega) \left[ f_{v}(\vec{k} - \lambda_{h,b}\vec{Q})  \right. \right. \nonumber \\
 & & \left. \left. - f_{c}(\vec{k} + \lambda_{e,b}\vec{Q}) \right] -  f_{c}(\vec{k} + \lambda_{e,b}\vec{Q}) \left[ 1 -  f_{v}(\vec{k} - \lambda_{h,b}\vec{Q}) \right] \right\} \nonumber \\
\end{eqnarray}
$f_{c/v}$ are the layer $b$ conduction and valence band electron occupation factors, and $\chi_{\vec{k},b}(\vec{Q})$ is related to the interband momentum matrix element in layer $b$ by the expression,
\begin{eqnarray}
 \chi_{\vec{k},b}(\vec{Q}) & = & \vec{P}_{vc,s}(\vec{k}-\lambda_{h}\vec{Q},\vec{k}+\lambda_{e}\vec{Q}).\hat{x} \nonumber \\
& & \times e^{i\left[ \tau \phi_{\vec{k}+\lambda_{e}\vec{Q}} + \tau \phi_{\vec{k}-\lambda_{h}\vec{Q}} \right]} \label{eq:chi2}
\end{eqnarray}
We define the momentum $k^{*}$ and the energy $E^{*}$ by the energy conservation relations, $E_{ex,L/T,a,\alpha}(Q) = E_{g,b} + E^{*} + \hbar^{2}Q^{2}/2m_{ex,b}$ and $E^{*} = \hbar^{2}(k^{*})^{2}/2m_{r,b}$. Given the number of possible final states (corresponding to different values of $\vec{k}$) and the fast electron and hole scattering rates in TMDs, we don't expect coherent oscillations. Assuming that the narrower bandgap material is in the ground state with a full valence band and an empty conduction band, the energy transfer times can be expressed as,
\begin{eqnarray}
  \frac{1}{\tau_{E,T,\alpha}} & \approx &   \frac{\pi \hbar}{2} g_{free,b}(E^{*}) \nonumber \\
  & & \times \left(2\eta_{o} \frac{e^{2}}{m^{2}_{o}} |\chi_{ex,a,\alpha}(0,\vec{Q})\chi_{\vec{k}^{*},b}(\vec{Q})| \right)^{2} \nonumber \\
  & & \times \frac{ e^{2i\sqrt{f_{\vec{Q},T,a,\alpha}} d/\hbar c}}{|f_{\vec{Q},T,a,\alpha}|} \label{eq:simple5} \\
  \frac{1}{\tau_{E,L,\alpha}} & \approx &   \frac{\pi \hbar}{2} g_{free,b}(E^{*}) \nonumber \\
& & \times \left(2\eta_{o} \frac{e^{2}}{m^{2}_{o}} |\chi_{ex,a,\alpha}(0,\vec{Q})\chi_{\vec{k}^{*},b}(\vec{Q})| \right)^{2} \nonumber \\
& & \times e^{2i\sqrt{f_{\vec{Q},L,a,\alpha}} d/\hbar c} \frac{|f_{\vec{Q},L,a,\alpha}|}{E^{4}_{ex,L,a,\alpha}(\vec{Q})} \label{eq:simple6}
\end{eqnarray}
where $f_{\vec{Q},L/T,a,\alpha} = (E^{2}_{ex,L/T,a,\alpha}(\vec{Q}) - \hbar^{2}Q^{2}c^{2})$, and $\sqrt{f_{\vec{Q},L/T,a,\alpha}} = i\sqrt{|f_{\vec{Q},L/T,a,\alpha}|}$ when $f_{\vec{Q},L/T,a,\alpha}<0$ outside the light cone. Here, $g_{free,b}(E)$ is the joint density of states for the creation of free electron-hole pairs (per valley/spin) in layer $b$. Again note that the energy transfer times do not depend on the scattering/dephasing rate. The expressions above differ from the corresponding expressions for free excitons, given earlier in (\ref{eq:simple1}) and (\ref{eq:simple2}), by a multiplicative factor of  $\pi \gamma g_{free,b}(E^{*}) |\chi_{\vec{k}^{*},b}(\vec{Q}) /\chi_{ex,a,\alpha}(0,\vec{Q})|^{2}$. This factor is expected to be much smaller than unity for most TMD pairs. Still, the energy transfer times for longitudinal excitons in layer $a$ can range from a picosecond to tens of picoseconds (depending on the initial exciton momentum $\vec{Q}$ and the magnitude of the joint density of states for free electron-hole pair creation in layer $b$) for interlayer spacings smaller than 10 nm.

As an example, we consider the case of MoS$_{2}$ and MoTe$_{2}$ layers. The exciton energy in MoS$_{2}$ is $\sim$1.9 eV and the quasiparticle bandgap $E_{g,b}$ of MoTe$_{2}$ is $\sim$1.7 eV. Assuming an exciton FWHM of 30 meV in MoS$_{2}$ and parameters of MoTe$_{2}$ as given in the literature~\cite{Yang15}, the value of $\pi \gamma g_{fee,b}(E^{*}) |\chi_{\vec{k}^{*},b}(\vec{Q}) /\chi_{ex,a,1s}(0,\vec{Q})|^{2}$ is found to be in the 0.07-0.08 range (for different momenta of the initial exciton state in MoS$_{2}$). Therefore, the energy transfer times between ($1s$) excitons in MoS$_{2}$ and free electron-hole pairs in MoTe$_{2}$ will be approximately 12-14 times those given in Figure~\ref{fig:fig3} for the case of $1s$ excitons in two identical MoS$_{2}$ layers.

\section{Discussion and Conclusion}
In this paper we presented results on the energy transfer rates between excitons in 2D TMD monolayers. The results show that the energy transfer rates can be very fast. Exciton energy transfer can potentially be used to design novel optoelectronic devices with TMD monolayers. To date, the authors are not aware of an experimental studies in this area. However, the results presented in this paper can be easily tested experimentally.  

The theory presented in this paper has certain limitations and care needs to exercised when interpreting the results and comparing these results with experiments:
\begin{enumerate}
\item The technique used in this paper is valid provided the exciton optical conductivity~\cite{Wang15c,Changjian14} $\sigma_{\alpha}(\omega)$ satisfies $|\eta_{o} \sigma_{\alpha}(\omega)| << 1$, which is typically the case in TMDs~\cite{Wang15c,Changjian14}. If $|\eta_{o} \sigma_{\alpha}(\omega)| \ge 1$, the vacuum field in the vicinity of the TMD layers will get modified and the expression for the field given in (\ref{eq:field}) will no longer be valid. The field would then need to be quantized in the presence of the TMD layers~\cite{Loudon95}, a task beyond the scope of this paper.
\item In plotting all the results, a momentum-independent exciton FWHM linewidth was used.  Exciton intralayer scattering and dephasing rates are expected to depend on the exciton momentum. Since the exciton energy transfer rates depend, in most cases, on the exciton scattering rates, a quantitative theory or experimental data for momentum-dependent exciton scattering rate in TMDs is needed for a better understanding of the dependence of the energy transfer rates on exciton momenta. 
\item It is well known that radiation emission and absorption rates are affected by the presence of dielectric interfaces~\cite{Loudon91}. Most experiments on TMD layers are performed with the layers placed on dielectric substrates. The influence of nearby dielectrics would need to be taken into account in comparing theory with experiments.   
\end{enumerate}

\section{Acknowledgments}
The authors would like to acknowledge helpful discussions with Jared Strait, Paul L. McEuen and Michael G. Spencer, and support from CCMR under NSF grant number DMR-1120296, AFOSR-MURI under grant number FA9550-09-1-0705, and ONR under grant number N00014-12-1-0072.

\section{Appendix}  \label{app1}

\subsection{Exciton Self-Energies in a Single TMD Layer}
In this section, we calculate the exciton self-energy in TMD monolayers. For the sake of simplicity we will assume that there is only one significant exciton level labeled by $\alpha$. The non-interacting (bare) retarded Green's function for the exciton field is~\cite{Mahan00},
\begin{equation}
  G^{oR}_{\vec{Q},L/T,\alpha}(t-t') = -\frac{i}{\hbar}\theta(t-t')\langle[C_{\vec{Q},L/T,\alpha}(t),C_{-\vec{Q},L/T,\alpha}(t')]\rangle
  \end{equation}
The exciton operator $C_{\vec{Q},L/T,\alpha}$ was defined earlier in (\ref{eq:fields}). In the Fourier domain the Green's function is,
\begin{eqnarray}
 G^{oR}_{\vec{Q},L/T,\alpha}(\omega) & =  & \frac{2E_{ex,\alpha}(\vec{Q})}{(\hbar\omega)^{2}-E_{ex,\alpha}(\vec{Q})^{2} + i\eta}
\end{eqnarray}
The non-interacting (bare) retarded radiation Green's function is defined as,
\begin{eqnarray}
D^{oR}_{\vec{q},L/T}(t-t') & = & -\frac{i}{\hbar}\theta(t-t') \, \langle[ R_{\vec{q},L/T}(t),R_{-\vec{q},L/T}(t') ]\rangle \nonumber \\
D^{oR}_{\vec{q},L/T}(\omega) & = &    \frac{2\hbar \omega_{q}}{(\hbar\omega)^{2} - (\hbar \omega_{q})^{2} + i\eta} \nonumber \\
\end{eqnarray}
First, we dress the radiation Green's function with term $H'_{int}$ in the exciton-photon interaction Hamiltonian that is quadratic in the vector potential (see Section (\ref{sec:quad}). The resulting dressed Green's functions $D^{R}_{\vec{q},L/T}(\omega)$ can be expressed in a form that will be useful later,
\begin{eqnarray}
  & & \int \frac{dq_{z}}{2\pi} \, \frac{\hbar}{\epsilon_{o} \omega_{q}} D^{R}_{\vec{q},T}(\omega) = \nonumber \\
  & & \frac{\int \frac{dq_{z}}{2\pi} \frac{\hbar}{\epsilon_{o} \omega_{q}} D^{oR}_{\vec{q},T}(\omega)}{1 - 2\frac{e^{2}}{m^{2}_{o}} \frac{|\chi_{ex}(0,\vec{q}_{\parallel})|^{2}}{E_{ex,\alpha}(\vec{q}_{\parallel})} \int \frac{dq_{z}}{2\pi} \frac{\hbar}{\epsilon_{o} \omega_{q}} \, D^{oR}_{\vec{q},T}(\omega) } \\
  & &  \int \frac{dq_{z}}{2\pi} \, \frac{\hbar}{\epsilon_{o} \omega_{q}} \, \frac{|q_{z}|^{2}}{q^{2}} D^{R}_{\vec{q},L}(\omega) = \nonumber \\
  & & \frac{\int \frac{dq_{z}}{2\pi} \, \frac{\hbar}{\epsilon_{o} \omega_{q}} \, \frac{|q_{z}|^{2}}{q^{2}} D^{oR}_{\vec{q},L}(\omega)}{1 - 2\frac{e^{2}}{m^{2}_{o}} \frac{|\chi_{ex}(0,\vec{q}_{\parallel})|^{2}}{E_{ex,\alpha}(\vec{q}_{\parallel})} \int \frac{dq_{z}}{2\pi} \frac{\hbar}{\epsilon_{o} \omega_{q}} \, \frac{|q_{z}|^{2}}{q^{2}} D^{oR}_{\vec{q},L}(\omega) } 
\end{eqnarray}
The denominator on the right hand side in the above equations is generally small and may be neglected. But it plays an important role when off-shell exciton self-energies are desired, as shown below. The Green's functions for the radiation field are gauge-dependent. It is convenient to choose the temporal gauge in which the scalar potential is set equal to zero (and need not be taken into account separately)~\cite{Guad80}. Henceforth, all results will be given for the temporal gauge. The Dyson equation for the exciton Green's function is,
\begin{eqnarray}
& &   G^{R}_{\vec{q}_{\parallel},L/T,\alpha}(\omega) = \nonumber \\
  & & G^{oR}_{\vec{q}_{\parallel},L/T,\alpha}(\omega) \left[ 1 +  \Sigma^{R}_{\vec{q}_{\parallel},L/T,\alpha}(\omega) \, \cdotp \, G^{R}_{\vec{q}_{\parallel},L/T,\alpha}(\omega) \right]
  \end{eqnarray}
where the retarded self-energies are found to be,
\begin{eqnarray}
  \Sigma^{R}_{\vec{q}_{\parallel},T,\alpha}(\omega) & = & \frac{e^{2}}{m^{2}_{o}}  |\chi_{ex}(0,\vec{q}_{\parallel})|^{2} \int \frac{dq_{z}}{2\pi} \frac{\hbar}{\epsilon_{o} \omega_q} D^{R}_{\vec{q},T}(\omega) \nonumber \\
  \Sigma^{R}_{\vec{q}_{\parallel},L,\alpha}(\omega) & = & \frac{e^{2}}{m^{2}_{o}}  |\chi_{ex}(0,\vec{q}_{\parallel})|^{2} \nonumber \\
  & & \times \int \frac{dq_{z}}{2\pi} \frac{\hbar}{\epsilon_{o} \omega_q} \, \frac{(\omega^{2} - q_{\parallel}^{2}c^{2})}{\omega^{2}} D^{R}_{\vec{q},L}(\omega) \nonumber \\
  \label{eq:self}
\end{eqnarray}
The dressed exciton Green's functions become,
\begin{eqnarray}
  & & G^{R}_{\vec{q}_{\parallel},L/T,\alpha}(\omega) = \nonumber \\
  & & \frac{2E_{ex,\alpha}(\vec{q}_{\parallel})}{(\hbar\omega)^{2}-E_{ex,\alpha}(\vec{q}_{\parallel})^{2} - 2E_{ex,\alpha}(\vec{q}_{\parallel}) \Sigma^{R}_{\vec{q}_{\parallel},L/T,\alpha}(\omega)} \nonumber \\
  & = & \frac{2E_{ex,\alpha}(\vec{q}_{\parallel})  \left[ 1-  2 \, \Sigma^{oR}_{\vec{q}_{\parallel},L/T,\alpha}(\omega)/E_{ex,\alpha}(\vec{q}_{\parallel}) \right]        }{(\hbar\omega)^{2}-E_{ex,\alpha}(\vec{q}_{\parallel})^{2} - 2\frac{(\hbar \omega)^{2}}{E_{ex,\alpha}(\vec{q}_{\parallel})} \Sigma^{oR}_{\vec{q}_{\parallel},L/T,\alpha}(\omega)} \nonumber \\
  & \approx & \frac{2E_{ex,\alpha}(\vec{q}_{\parallel})}{(\hbar\omega)^{2}-E_{ex,\alpha}(\vec{q}_{\parallel})^{2} - 2\frac{(\hbar \omega)^{2}}{E_{ex,\alpha}(\vec{q}_{\parallel})} \Sigma^{oR}_{\vec{q}_{\parallel},L/T,\alpha}(\omega)} \nonumber \\
\end{eqnarray}  
Here, $\Sigma^{oR}_{\vec{q}_{\parallel},L/T,\alpha}(\omega)$ is the same as $\Sigma^{R}_{\vec{q}_{\parallel},L/T,\alpha}(\omega)$ given in (\ref{eq:self}) above except that the bare radiation Green's function $D^{oR}_{\vec{q},L/T}(\omega)$ is used in place of $D^{R}_{\vec{q},L/T}(\omega)$. Finally the exciton spectral density function $ A_{\vec{q}_{\parallel},L/T,\alpha}(\omega)$ can be related to the retarded Green's function as follows,
\begin{equation}
-2 \hbar \, {\rm Imag}\left\{  G^{R}_{\vec{q}_{\parallel},L/T,\alpha}(\omega) \right\} = A_{\vec{q}_{\parallel},L/T,\alpha}(\omega) - A_{\vec{q}_{\parallel},L/T,\alpha}(-\omega)
\end{equation}

\begin{figure}
  \begin{center}
   \epsfig{file=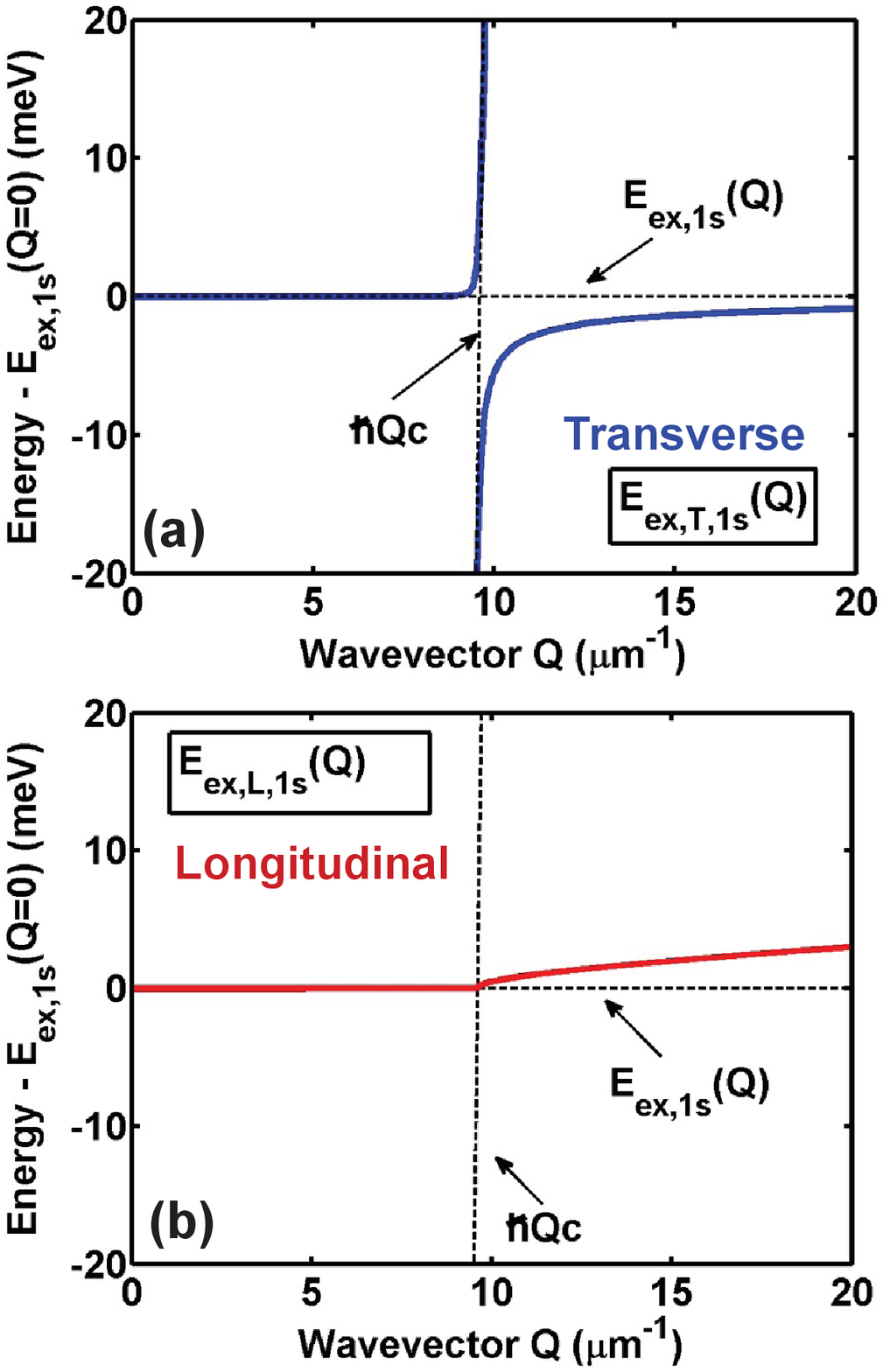,angle=0,width=0.4\textwidth}
    \caption{Calculated transverse and longitudinal exciton energy dispersions for the lowest energy ($1s$) exciton state in a suspended MoS$_{2}$ monolayer are plotted as a function of the in-plane momentum $Q$. Also shown are the photon and the bare exciton dispersion relations (dashed lines). (a) The transverse exciton energy dispersion consists of two curves. The upper curve has all the spectral weight for $Q << Q_{o}$ and the lower curve gets all the  weight when $Q >> Q_{o}$, and the spectral weight shifts from the upper curve to the lower curve around $Q \approx Q_{o}$. (b) The dispersion for the longitudinal exciton consists of only a single curve and the energy dispersion is linear in $Q$ for $Q >> Q_{o}$.}
    \label{fig:FigA1}
  \end{center}
\end{figure}

When $\omega > q_{\parallel}c$ (inside the light cone), both the self-energies $\Sigma^{R}_{\vec{q}_{\parallel},L/T,\alpha}(\omega)$ have a vanishingly small real part and a large magnitude of the imaginary part. The latter corresponds to the radiative lifetime of the exciton. The situation is reversed when $\omega < q_{\parallel}c$ and then the magnitude of the real part of the self-energy becomes large and the imaginary part vanishes. Therefore, when $E_{ex,\alpha}(\vec{q}_{\parallel}) > \hbar q_{\parallel}c$ (inside the light cone) one can ignore corrections to the exciton dispersion, and the radiative lifetime of the exciton can be related to the imaginary part of the self-energy evaluated on the shell,
\begin{equation}
  \frac{1}{\tau_{sp,\vec{q}_{\parallel},L/T,\alpha}} = -\frac{2}{\hbar}{\rm Imag}\left\{ \Sigma^{R}_{\vec{q}_{\parallel},L/T,\alpha}(\omega) \right\}_{\hbar \omega = E_{ex,\alpha}(\vec{Q})}
\end{equation}
and, we get~\cite{Wang16,Gartstein15},
\begin{eqnarray}
  \frac{1}{\tau_{sp,\vec{q}_{\parallel},T,\alpha}} & = & \frac{ 2\eta_{o} e^{2}}{m^{2}_{o}} |\chi_{ex}(0,\vec{q}_{\parallel})|^{2}\frac{1}{\sqrt{E^{2}_{ex,\alpha}(\vec{q}_{\parallel}) - (\hbar q_{\parallel} c)^{2}}} \nonumber \\
  \frac{1}{\tau_{sp,\vec{q}_{\parallel},L,\alpha}} & = &  \frac{ 2\eta_{o} e^{2}}{m^{2}_{o}} |\chi_{ex}(0,\vec{q}_{\parallel})|^{2}\frac{\sqrt{E^{2}_{ex,\alpha}(\vec{q}_{\parallel}) - (\hbar q_{\parallel} c)^{2}}}{E^{2}_{ex,\alpha}(\vec{q}_{\parallel})} \nonumber \\
\end{eqnarray}
When $\omega < q_{\parallel}c$ and the self-energies are real, we get~\cite{Wang16,Gartstein15},
\begin{eqnarray}
  \Sigma^{R}_{\vec{q}_{\parallel},T,\alpha}(\omega) & = & - \frac{\eta_{o} \hbar^{2} e^{2}}{m^{2}_{o}} \frac{|\chi_{ex}(0,\vec{q}_{\parallel})|^{2}}{E^{2}_{ex,\alpha}(\vec{q}_{\parallel})}  \frac{\omega^{2}}{\sqrt{(q_{\parallel} c)^{2} - \omega^{2}}} \nonumber \\
  \Sigma^{R}_{\vec{q}_{\parallel},L,\alpha}(\omega) & = &  \frac{\eta_{o} \hbar^{2} e^{2}}{m^{2}_{o}} \frac{|\chi_{ex}(0,\vec{q}_{\parallel})|^{2}}{E^{2}_{ex,\alpha}(\vec{q}_{\parallel})} \sqrt{ (q_{\parallel} c)^{2} - \omega^{2}} \nonumber \\
\end{eqnarray}
Note that the corrections obtained from $H'_{int}$, which was quadratic in the vector potential, are important for obtaining the correct expressions for the exciton self-energies far off-shell. 

\subsection{Exciton Energy Dispersions in a Single TMD Layer}
The exciton self-energy expressions given above can be used to obtain energy dispersions, $E_{ex,L/T,\alpha}(\vec{q}_{\parallel})$, for the longitudinal and transverse excitons. The results are shown in Figure ~\ref{fig:FigA1} for the lowest energy ($1s$) exciton state in a suspended MoS$_{2}$ monolayer. We define $Q_{o}$ by the equation $E_{ex,1s}(Q_{o}) = \hbar Q_{o} c$. The dispersion for the transverse exciton consists of two curves (corresponding to the distinct poles of the Green's function). The upper curve has all the spectral weight for $Q << Q_{o}$ and the lower curve gets all the  weight when $Q >> Q_{o}$, and the spectral weight shifts from the upper curve to the lower curve around $Q \approx Q_{o}$. The dispersion for the longitudinal exciton consists of only a single curve and the energy dispersion is linear in $Q$ for $Q >> Q_{o}$. For $Q > Q_{o}$, $E_{ex,L,1s}(\vec{Q}) > E_{ex,T,1s}(\vec{Q})$. Therefore, in a thermal ensemble of excitons, transverse exciton density is expected to exceed the longitudinal exciton density. Note that there are negligibly small corrections to the exciton dispersion within the light cone for both the transverse and longitudinal excitons.


\begin{thebibliography}{99}

\bibitem{fai10} K. F. Mak, C. Lee, J. Hone, J. Shan, and T. F. Heinz, Phys. Rev. Lett. 105, 136805 (2010).
\bibitem{fai12} K. F. Mak, K. He, J. Shan, and T. F. Heinz, Nat. Nanotech. 7, 494 (2012).
\bibitem{xu13}  J. S. Ross, S. Wu, H. Yu, N. J. Ghimire, A. M. Jones, G. Aivazian, J. Yan, D. G. Mandrus, D. Xiao, W. Yao, and X. Xu, Nat. Comm. 4, 1474 (2013). 	
\bibitem{timothy} T. C. Berkelbach, M. S. Hybertsen, and D. R. Reichman, Phys. Rev. B 88, 045318 (2013).
\bibitem{Chernikov14} A. Chernikov, T. C. Berkelbach, H. M. Hill, A. Rigosi, Y. Li, O. B. Aslan, D. R. Reichman, M. S. Hybertsen, T. F. Heinz, Phys. rev. Lett., 113, 076802 (2014). 
\bibitem{Changjian14} C. Zhang, H. Wang, W. Chan, C. Manolatou, F. Rana, Phys. Rev. B, 89, 205436 (2014). 
\bibitem{Wang16} H. Wang, C. Zhang, W. Chan, C. Manolatou, S. Tiwari, F. Rana, Phys. Rev. B, 93, 045407 (2016). 
\bibitem{Menon14} X. Liu, T. Galfsky, Z. Sun, F. Xia, E. Lin, Y. Lee, S. Kéna-Cohen, V. M. Menon, Nature Photonics, 9, 30 (2015).
\bibitem{Vasil15} M. I. Vasilevskiy, D. G. Santiago-Pérez, C. Trallero-Giner, N. M. R. Peres, A. Kavokin, Phys. Rev. B 92, 245435 (2015). \bibitem{Gartstein15} Y. N. Gartstein, X. Li, C. Zhang, Phys. Rev. B, 92, 075445 (2015). 
\bibitem{Moody15} G. Moody, C. K. Dass, K. Hao, C.-H. Chen, L.-J. Li, A. Singh, K. Tran, G. Clark, X. Xu, G. BerghÃ¤user, E. Malic, A. Knorr, X. Li, Nature Communications, 6, 8315 (2015).
\bibitem{Huber15} C. Poellmann, P. Steinleitner, U. Leierseder, P. Nagler, G. Plechinger, M. Porer, R. Bratschitsch, C. SchÃ¼ller, T. Korn, R. Huber, Nature Materials, 14, 889 (2015).
\bibitem{Marie15} X. Marie, B. Urbaszek, Nature Materials, 14, 860 (2015). 
\bibitem{Javey14} H. Fang, C. Battaglia, C. Carraro, S. Nemsak, B. Ozdol, J. Seuk Kang, H. A. Bechtel, S. B. Desai, F. Kronast, A. A. Unal, G. Conti, C. Conlon, G. K. Palsson, M. C. Martin, A. M. Minor, C. S. Fadley, E. Yablonovitch, R. Maboudian, A. Javey, PNAS, 111, 6198 (2014). 
\bibitem{Hong14} X. Hong, J. Kim, S. Shi, Y. Zhang, C. Jin, Y. Sun, S. Tongay, J. Wu, Y. Zhang, F. Wang, Nature Nanotechnology, 9, 682–686 (2014).
\bibitem{Rigosi15} A. F. Rigosi, H. M. Hill, Y. Li, A. Chernikov, and T. F. Heinz, Nano Letters, 15, 5033 (2015).  	
\bibitem{Wang15b} H. Wang, J. H. Strait, C. Zhang, W. Chen, C. Manolatou, S. Tiwari, F. Rana, Phys. Rev. B 91, 165411 (2015). 
\bibitem{Lam12} T. Cheiwchanchamnangij and W. R. L. Lambrecht, Phys. Rev. B 85, 205302 (2012).
\bibitem{Falko13} A. Kormanyos, V. Zolyomi, N. D. Drummond, P. Rakyta, G. Burkard and V. I. Falko, Phys. Rev. B 88, 045416 (2013). 
\bibitem{yao12} D. Xiao, Gui-Bin Liu, W. Feng, X. Xu, and W. Yao, Phys. Rev. Lett. 108, 196802 (2012).
\bibitem{Girlanda95} S. Savasta, R. Girlanda, Solid State Communications, 96, 517 (1995). 
\bibitem{haugbook} H. Haug, S. W. Koch, {\em Quantum Theory of the Optical and Electronic Properties of Semiconductors}, World Scientific Publishing, Singapore (1990). 
\bibitem{Mahan00} G. D. Mahan, {\em Many Particle Physics}, Springer, NY (2000).
\bibitem{PandA89} C. Cohen-Tannoudji, J. Dupont-Roc, G. Grynberg, {\em Photons and Atoms: Introduction to Quantum Electrodynamics}, Wiley, NY (1989).  
\bibitem{Andrews04} D. L. Andrews, D. S. Bradshaw, European Journal of Physics, 25, 845 (2004). 
\bibitem{Forster48} T. Forster, Annals of Physics, 437, 55 (1948).  
\bibitem{Tomita96} A. Tomita, J. Shah, Phys. Rev. B, 53, 10793 (1996). 
\bibitem{Lyo00} S. K. Lyo, Phys. Rev. B, 62, 13641 (2000).    
\bibitem{Wang15c} H. Wang, C. Zhang, F. Rana, Nano Letters, 15, 339 (2015). 
\bibitem{Guad80} E. Guadagnini, Il Nuovo Cimento, 57A, 294 (1980). 
\bibitem{Yang15} J. Yang, T. Lu, Y. W. Myint, J. Pei, D. Macdonald, J. Zheng, Y. Lu, ACS Nano, 9, 6603 (2015). 
\bibitem{Loudon95} R. Matloob, R. Loudon, S. M. Barnett, J. Jeffers, Phys. Rev. A 52, 4823 (1995). 
\bibitem{Loudon91} H. Khosravi, R. Loudon, Proc. R. Soc. London A, 433, 337 (1991).  
\bibitem{Citrin94} D. S. Citrin, Phys. Rev. B, 49, 1943 (1994). 
  
\end{thebibliography}
\end{document}